\documentstyle [12pt,a4p,epsfig,amsmath,multicol]{article}
\begin{document}
\vspace*{0.6cm}
\begin{center}
{\bf Why Einstein, Podolsky and Rosen did not prove that quantum mechanics is `incomplete'}
\end{center}
\vspace*{0.6cm}
\centerline{\footnotesize J.H.Field}
\baselineskip=13pt
\centerline{\footnotesize\it D\'{e}partement de Physique Nucl\'{e}aire et 
 Corpusculaire, Universit\'{e} de Gen\`{e}ve}
\baselineskip=12pt
\centerline{\footnotesize\it 24, quai Ernest-Ansermet CH-1211 Gen\`{e}ve 4. }
\centerline{\footnotesize E-mail: john.field@cern.ch}
\baselineskip=13pt
 
\vspace*{0.9cm}
\abstract{ It is shown that the Einstein-Podolsky-Rosen conclusion
 concerning the `incompleteness' of Quantum Mechanics is invalidated by
 two logical errors in their argument. If it were possible to 
 perform the proposed gedanken experiment it would, in fact, show that  Quantum 
 Mechanics is `complete' for the observables discussed.
 Because, however, of the non square-integrable nature of the 
 wave function, the proposed experiment gives vanishing
 probabilities for measurements performed in finite intervals 
 of configuration or momentum space. Hence no conclusion as
 to the `completeness', or otherwise, of Quantum Mechanics can
 be drawn from the experiment.}
\vspace*{0.9cm}
\normalsize\baselineskip=15pt
\setcounter{footnote}{0}
\renewcommand{\thefootnote}{\alph{footnote}}

\vspace*{0.4cm}

\newpage
\par Perhaps no other paper written in the 20th Century generated
as much debate about questions related to the foundations of physics and 
their philosophical implications than that of Einstein, Podolsky and
Rosen (EPR)~\cite{x1}. However, after the initial replies written by
Bohr~\cite{x2}, Furry~\cite{x3} and Schr\"{o}dinger~\cite{x4}, there
has been very little critical discussion of the EPR paper itself in
the literature\footnote{Almost all subsequent discussion of `EPR experiments'
in the literature is, instead, based on Bohm's gedanken
experiment ~\cite{x5} involving correlated spin measurements.}.
 In this article a reappraisal of the EPR paper
 is made and the following conclusions are drawn:
\begin{itemize}
\item[(i)] The argument presented by EPR to demonstrate the 
`incompleteness' of Quantum Mechanics (QM) is invalidated by two logical
 errors.
\item[(ii)] The gedanken experiment proposed by EPR cannot be carried
 out if the usual probabilistic interpretation of QM is correct, and
 so no physical conclusions can be drawn from the experiment.
\end{itemize}
\par Following EPR, a theory is said to give a `complete' description
of a physical quantity if the following condition is satisfied:
\par `Without, in any way, disturbing a system, we can predict with
 certainty (i.e. with probability equal to unity) the value of the 
 physical quantity'.
\par If this is the case, EPR associate an `Element of Physical Reality'
to the the corresponding quantity. EPR also require that, in a `complete'
theory:
\par `Every Element of Physical Reality must have a counterpart
 in the physical theory'.
\par This hypothesis is not particularly important since it must
  necessarily be true if the theory is able to predict the value of
 the corresponding physical quantity. 
\par The EPR gedanken experiment will first be discussed from a 
purely logical viewpoint. Secondly, the conceptual feasiblity, within
QM, of the proposed experiment is examined. The following hypotheses
 are defined:
\begin{itemize}
\item  $QMT$~: QM is a true theory within its domain of applicablity.
\item $QMTC(A,B,..)$~: QM is a true, complete, theory for the physical
 quantities A,B,.. .
\item $PRNC(A,B)$~: Elements of Physical Reality exist for each of a 
      pair of physical quantities A, B  with non-commuting operators
      in QM.
\end{itemize}
 The EPR gedanken experiment is based solely on the hypothesis $QMT$
 (Quantum Mechanics True). Contrary to the statement of EPR, it is
 {\it not necessary} to assume, at the outset, that QM is also a
complete theory (hypothesis $QMTC$). This is EPR's first logical error.
  In fact, applying $QMT$ and assuming
also that a quantum mechanical system of two correlated particles with
a certain well-defined wave function can be constructed, EPR found 
that Elements of Physical Reality 
 apparently {\it can} be assigned to each 
of the quantities $P$ and $Q$ that that have non-commuting operators.
 EPR thus found that the proposition $PRNC(P,Q)$ follows from $QMT$ alone
 according to their interpretation of the results of the gedanken experiment.
 After correction\footnote{i.e. replacing in the statement of EPR the
 hypothesis $QMTC$ by $QMT$.}, the final statement of the result of the
 gedanken experiment is:
 \par `Starting from the assumption of the correctness of QM
 (i.e. hypothesis $QMT$) we arrived at the conclusion that two physical
  quantities with non-commuting observables can have simultaneous
  reality.'
\par According to EPR's definitions,
if two physical quantities have corresponding elements of physical
 reality, then the theory is a complete one for these quantities.
 In symbols\footnote{Each hypothesis is assumed to be either true or false. The
symbols $\otimes$ and $\oplus$ denote, respectively logical `and'
 and  `exclusive or'. The latter is defined in such a way  that if $X \oplus Y = TRUE$. the 
  possiblities that $X$ and $Y$ are both $TRUE$, or both $FALSE$, are excluded. A bar
  on a logical variable indicates negation.}:
\begin{equation}
 PRNC(P,Q) \otimes QMTC(P,Q) = TRUE 
\end{equation}
Using De Morgan's Theorem, (1) implies:
\begin{equation}
\overline{PRNC(P,Q)} \oplus \overline{QMTC(P,Q)} = FALSE 
\end{equation}
 where in (2) and the following the symbol `$\oplus$' denotes an `exclusive or'
 \footnote{The meaning of the `exclusive or' $X \oplus Y = FALSE$ is that either
  $X$ and $Y$ are both true, or they are both false.}.
However, as will be seen below, EPR state that the right side of (2) is TRUE from which it follows
 instead of (1), that:
\begin{equation}
 PRNC(P,Q) \otimes QMTC(P,Q) = FALSE 
\end{equation}
Since the gedanken experiment shows that:
\[ PRNC(P,Q)=TRUE, \]
 the erroneous conclusion is drawn, on the basis of (3),  that:
\[ QMTC(P,Q)=FALSE. \]
i.e. that QM is an incomplete theory. This was EPR's final conclusion. It was
 actually reached using a {\it reductio ad absurdum} argument that is reviewed below.
 The basic assertion of EPR (actually, as shown above, in contradiction
to the result of their gedanken experiment) is the negation of the proposition (2):
\begin{equation}
\overline{PRNC(P,Q)}\oplus\overline{QMTC(P,Q)} = TRUE 
\end{equation} 
 This is EPR's second logical error. 
How is this assertion justified in the EPR paper? After discussion of
quantum mechanical measurements on a {\it single particle}, with no
obvious relevance to the case of {\it two correlated particles}
as used in their gedanken experiment, EPR state that:
\par  `From this it follows [1]{\it the quantum mechanical description
 of reality given by the wave function is not complete or} [2]{\it when the
 operators corresponding to two physical quantities do not commute the 
 two quantities cannot have simultaneous reality.}  For if both of
 them had simultaneous reality - and thus definite values - these values
 would enter into the complete description according to the condition
 of completeness. If the wave function provided such a complete 
 description of reality it would contain these values, these would be 
 predictable. This not being the case we are left with the
 alternatives stated.'
 \par (Italics in the original)
 \par The italicised statement, expressed symbolically by Eqn.(4),
  seems to be justified
 by the preceding discussion in the paper of non commuting observables
 (position and momentum) for a {\it single } particle, not the correlated
 two particle system of the gedanken experiment subsequently presented.
 In fact no justification is given by EPR for the application,
 {\it a priori}, of propositions [1] and [2] and their relation Eqn.(4)
  to the gedanken experiment.
 Even so, one can still ask what is the meaning of EPR's assertion in
 Eqn.(4)?  As quoted above, EPR carefully explain that the negation of the proposition
 [2] implies that the quantum mechanical description of two 
 commuting observables is complete, i.e.
\[   \overline{[2]} \equiv PRNC(A,B) \Rightarrow QMTC(A,B) \]
  or, equivalently,
 \[   [2] \equiv \overline{PRNC(A,B)} \Rightarrow \overline{QMTC(A,B)} \]
 where the symbol  $\Rightarrow$ is used for  `logically implies', i.e. $X \Rightarrow Y$
 means that if $X$ is $TRUE(FALSE)$ then $Y$ is $TRUE(FALSE)$ and {\it vice versa}.
 But just this condition i.e. both $\overline{PRNC(A,B)} = TRUE$ and 
 $\overline{QMTC(A,B)} = TRUE$ is one of the two possibilities that are excluded
 by the definition of the `exclusive or' proposition (4) that EPR assume to be correct!
 \par To summarise, the EPR argument is based on the `exclusive or' proposition:
   \[ X \oplus Y = TRUE \]
  which implies that the only possiblities are: $X = FALSE$ and $Y = TRUE$ or {\it vice versa},
  the cases when $X$ and $Y$ are both true or both false being excluded.
    But the truth (or falsehood) of the proposition
      $X \equiv \overline{PRNC(P,Q)}$ entails the truth (or falsehood) of
        the proposition
      $Y \equiv \overline{QMTC(P,Q)}$. Therefore if $X$ is false ---the 
     claimed conclusion of EPR's analysis of their gedanken experiment---
      $Y$ must also be false, so that QM is then a complete theory,
     not an incomplete one as claimed by EPR. In fact the true conclusion,
    that both $X$ and $Y$ are false, contradicts EPR´s initial proposition (4).
     As pointed out above, when 
     $X$ and $Y$ are both false the `$TRUE$' on the right side of EPR's initial
     proposition must be replaced by `$FALSE$'. 
     Indeed, $\overline{PRNC(P,Q)}$ \newline---quantum mechanics is incomplete for
     the non-commuting
      variables $P$,$Q$--- is a special case implied by the more general proposition 
      $\overline{QMTC(A,B)}$ for arbitary non-commuting variables $A$ and $B$, when $A = P$ and $B = Q$.  
      \par An example of an absurd (self-contradictory) conclusion that is obtained
          from asserting the truth of an exclusive or proposition, when both
          of the related propositions is true, is the following. Assume such a proposition
        is true when $X =$ `Aristotle was mortal' and $Y=$  `Aristotle was a man'.
        Men are a subset of all mortal beings just as the `incomplete' variables in
         $\overline{PRNC(P,Q)}$ are a subset of those in  $\overline{QMTC(A,B)}$.
         Since Aristotle died, $X$ is true. It follows then from the `exclusive or'
         that Aristotle was not a man, in contradiction to the inital proposition $Y$!
    \par Since EPR were assuming the correctness of the `exclusive or' (4) and the 
        gedanken experiment is claimed to show that\footnote{Since the arguments,
        $P$ and $Q$, of $PRNC$ and $QMTC$ are the same they are omitted, 
        for brevity, in the following}  $PRNC = TRUE$ or
        $\overline{PRNC} = FALSE$, they could have immediately concluded from 
         (4) that  $\overline{QMTC} = TRUE$ ---quantum mechanics is not complete. 
         Actually, however, they reached the same conclusion, based on (4), by a more 
        convoluted  {\it reductio ad absurdum} argument. EPR introduced, as well as
        (4), the initial hypothesis: $QMTC = TRUE$. They then claimed that the
        result of the gedanken experiment which was $\overline{PRNC} = FALSE$ and hence
        also $\overline{QMTC} = FALSE$ was a logical consequence of the assumption
         $QMTC = TRUE$. In spite of the fact that $\overline{QMTC} = FALSE$
         is logically equivalent to the claimed initial proposition ~$QMTC = TRUE$, EPR noted 
        that the result of the gedanken experiment,
        $\overline{PRNC} = FALSE$, together with
        the proposition (4),
        the correctness of which they assume, implies that $\overline{QMTC} = TRUE$, in
        contradiction with their initial proposition  $QMTC = TRUE$. EPR then deduce from
         this contradiction,
        by {\it reductio ad absurdum}, that the initial propostion $QMTC = TRUE$ must
        be false, so that quantum mechanics is incomplete. In fact the result of the
        gedanken experiment is consistent with the assumption $QMTC = TRUE$, it
        is the other initial proposition,(4), that must be rejected as erroneous.
          Indeed the proposition (4) is erroneous ---self contradictory--- given only
       the meanings of the two related propositions. 
         The {\it reductio} argument is therefore valid, but the intial hypothesis that must be 
         rejected is not $QMTC = TRUE$ but (4)! As stated above in any case the initial
         hypothesis of the gedanken experiment is $QMT$ {\it not} $QMTC$ and the result
         $QMTC = TRUE$ is derived from the gedanken experiment without, contrary to
          EPR's assertion, first assuming that $QMTC = TRUE$.  
 \par Correcting the logical errors described above, it might seem that the EPR 
 experiment establishes the `completeness' of quantum mechanics for
 the two non-commuting quantities P and Q. For this, however, it is
necessary that the suggested gedanken experiment
  can, at least in principle,
 be performed. It will now be shown that this is not the case, so 
 that no conclusion can be drawn as to the `completeness', or
 otherwise, of quantum mechanics, by the arguments presented by
 EPR.
\par The spatial wave function of the correlated two particle system
discussed by EPR is:
\begin{equation}
\Psi(x_1,x_2) = \int_{-\infty}^{\infty}dp \exp {\frac{2 \pi i}{h}(x_1-x_2+x_0)p}
  = h \delta(x_1-x_2+x_0)
\end{equation}
 The probability that the particle 1 will be observed in the interval
 $a < x_1 <b$, for any position of the particle 2, can be written as: 
\begin{eqnarray}
 P(a < x_1 <b) & = &  Lim (L \rightarrow \infty)
\frac{\int_{a}^{b} dx_1 \int_{-\infty}^{\infty} dx_2
|\Psi(x_1,x_2)|^2}
{\int_{-L}^{L} dx_1 \int_{-\infty}^{\infty} dx_2
|\Psi(x_1,x_2)|^2} \nonumber \\
 & = &  Lim (L \rightarrow \infty) \frac{b-a}{2L} = 0     
\end{eqnarray}
 The particle 1 cannot, therefore, be observed in any finite
 interval of $x_1$, and so the $Q$ measurement suggested in the EPR
 gedanken experiment cannot be carried out.
\par By making Fourier transforms with respect to $x_1$ and $x_2$
the momentum wavefunction corresponding to (5) is found to be:
\begin{equation}
\Psi(p_1,p_2) = \frac{h^2}{2 \pi} \exp \frac{2 \pi i p_1 x_0}{h}
 \delta(p_1+p_2)
\end{equation}
The probability to observe $p_1$ in the range $p_a <  p_1 < p_b$ for
 any value of $p_2$ is:
\begin{eqnarray}
 P(p_a < x_1 < p_b) & = &  Lim (p \rightarrow \infty)
\frac{\int_{p_a}^{p_b} dp_1 \int_{-\infty}^{\infty} dp_2
|\Psi(p_1,p_2)|^2}
{\int_{-p}^{p} dp_1 \int_{-\infty}^{\infty} dp_2
|\Psi(p_1,p_2)|^2} \nonumber \\
 & = &  Lim (p \rightarrow \infty) \frac{p_b-p_a}{2p} = 0     
\end{eqnarray}
 The momentum of particle 1 cannot be measured in any finite interval
 so that the proposed $p_2=P=-p_1$ measurement of the EPR gedanken
 experiment cannot be carried out. In fact, the correlated two particle
 wave function proposed by EPR is not square integrable either in 
 configuration or momentum space and so has no probabilistic 
 interpretation in QM. The single particle wavefunction discussed by
 EPR has the same shortcoming. Hence the `relative probability'
 $P(a,b)$ of EPR's Equation (6) also vanishes. While the statement
 that `all values of the coordinate are equally probable' is true,
 it is also true that the absolute probability to observe the
 particle in any finite coordinate interval is zero. 
\par The EPR two particle wavefunction is now modified to render it
 square integrable so that the results of the gedanken experiment may be
 interpreted according to the usual rules of QM. The suggested
`minimally modified' wavefunction is:
\begin{equation}
\tilde{\Psi}(x_1,x_2) = \frac{1}{(\sqrt{2\pi}\sigma_x)^{\frac{1}{2}}}
 \exp \left( {\frac{x_0^2-2x_1^2-2x_2^2}{16 \sigma_x^2}} \right)
 \delta(x_1-x_2+x_0)
\end{equation}
Like the EPR wavefunction (5), $\tilde{\Psi}$ vanishes unless $x_2=x_1+x_0$,
but it is square integrable and normalised:
\begin{equation}
\int_{-\infty}^{\infty}\int_{-\infty}^{\infty}|\tilde{\Psi}(x_1,x_2)|^2
dx_1dx_2 = 1
\end{equation}
The EPR wavefunction (5) is recovered in the limit $\sigma_x \rightarrow
\infty$. Performing a double Fourier transform on Eqn(9) yields the
 corresponding momentum wave function:
\begin{equation}
\tilde{\Psi}(p_1,p_2) = \frac{1}{\pi \sigma_p}
\exp \left( {-\frac{(p_1+p_2)^2}{2 \sigma_p^2}}\right)
 \exp \left(\frac{2 \pi i p_1 x_0}{h}\right) 
\end{equation}
 where
 \[ \sigma_p = h/4 \pi \sigma_x \]
 The wavefunction (11) is also square integrable and normalised:
\begin{equation}
\int_{-\infty}^{\infty}\int_{-\infty}^{\infty}|\tilde{\Psi}(p_1,p_2)|^2
dp_1dp_2 = 1
\end{equation}
and the EPR wavefunction (7) is recovered in the limit
 $\sigma_x \rightarrow \infty$,
 $\sigma_p \rightarrow 0$. Now, performing the EPR gedanken experiment,
using instead the wavefunctions (9) and (11) it becomes clear that it 
is no longer possible to associate `Elements of Physical Reality' to the
position Q and the momentum P of the second particle by performing
measurements on the first one. The probability $\delta P(x_1)$ that
the spatial position of the first particle lies in the interval
 $\delta x_1$ around $x_1$\footnote{i.e. that $x_1$ lies between $x_1-\delta x_1/2$ and
$x_1+\delta x_1/2$} is:
\begin{equation}
\delta P(x_1) = \frac{1}{\sqrt \pi \sigma_x} \exp \left(
-\frac{x_0^2}{8 \sigma_x^2}\right)
 \exp \left(-\frac{(2x_1+x_0)^2}{8 \sigma_x^2} \right) \delta x_1
\end{equation}
Because of the $\delta$-function in the wave function (9), this is
also the probability that $x_2$ lies in the interval of width
$\delta x_2 =\delta x_1$ around $x_2 = x_1 +x_0$. Measuring $x_1$ 
in the interval $\delta x_1$ then enables the certain prediction
that $x_2$ lies in the interval $\delta x_2$ around $x_2 = x_1 +x_0$.
However, to associate an `Element of Physical Reality' to $x_2$
requires that the
 {\it value} must be exactly predictable. For this it is necessary
that $\delta x_1 = \delta x_2 \rightarrow 0$. In this case $\delta P(x_1)$
vanishes and no possibility exists to measure the position of the
particle 1. The situation is then the same as in the case of the 
original EPR wavefunction (5). It is clear that, using the wavefunctions
 (9) and (10), the product of the uncertainties in P and Q can be much smaller
 than that required by the
Heisenberg uncertainty principle. However in order to thus determine Q, 
use is made of of the precise knowledge of the parameter $x_0$ of
the wavefunction, i.e. exact knowledge of how the wavefunction is 
prepared is required. But if {\it a priori} knowledge about
wavefunction preparation is admitted, it is trivial to show that
 observables with non-commuting operators can be simultaneously
 `known' with a joint precision far exceeding that allowed by the
momentum-space uncertainty relation. To give a concrete example of this,
the process of para-positronium annihilation at rest: $ e^+e^-
 \rightarrow \gamma \gamma$ may be considered. The uncertainty in
 the momentum 
$\Delta p$ of one of the decay photons is determined by the energy-time
 uncertainty relation\footnote{This is derived from the Fourier transform of the
 exponential decay law which gives a Breit-Wigner function of width $ \simeq c \Delta p$.}and the kinematical relation $c\Delta p = \Delta E$ to be: 
 \[\Delta p = \frac{h}{c \tau} \]
 where the mean lifetime of the decay process $\tau = 1.25 \times 10^{-10}$ sec. 
 The momentum-space uncertainty relation then predicts 
 \[ \Delta x > 3.75 {\rm cm} \]
 The technically simple measurement of the position of the photon
 in the direction parallel to its momentum to within
  $\pm1$mm (for example, by observing a recoil electron from Compton
 scattering of the photon~\cite{x9}) then allows simultaneous 
 knowledge of the position and momentum of the photon
 (whose quantum mechanical operators do not commute) with 
 an accuracy $\simeq$ 40 times better than `allowed' by the
 momentum-space uncertainty relation. Of course this uncertainty relation
 does indeed limit the precision of any attempt to 
 {\it simultaneously measure} a pair of non-commuting observables.
 However, as the counter example given above shows, it does not
 apply to {\it a priori} knowledge from state preparation,
 as used by EPR in the discussion of their gedanken experiment.
 There is therefore nothing remarkable (certainly no `paradox')
 in the fact that non-commuting observables can be `known' more 
 accurately than allowed by the uncertainty relation if information
 about state preparation is also included, as is the case for the
 EPR gedanken experiment. In fact, information from state preparation
 is essential for the EPR analysis of (hypothetical) measurements
  of the system described by the wavefunction (5). According to the
  latter the value of $x_2$ is fixed by a putative measurement of $x_1$ and {\it vice versa}.
   In both cases the information on the unmeasured variable is given by
   prior knowledge of the prepared wavefunction of the system.  
\par It has been stressed above, that no meaningful conclusions can be 
drawn from any  gedanken experiment based upon non square-integrable
wave functions. A similar criticism was made by Johansen
~\cite{x10} concerning a paper of Bell~\cite{x11} where the erroneous
conclusion was drawn, by the use of a non square-integrable
wave function, that states with a positive Wigner distribution
 (as is in fact the case for the EPR wave function (5)) necessarily
yield a local hidden variable model. A corollary is given by the
`complementary' limits discussed by Bohr~\cite{x2}, where an aspect
 of classical physics is recovered, yielding a precise position or
 momentum for a particle. Such exact limits are of limited physical
interest since the corresponding wavefunctions are not square 
 integrable for the conjugate variable, and so can have no
 physical interpretation within quantum mechanics. The Dirac
 $\delta$-function is a calculational device of extreme utility.
 It should never be forgotten, however, that it is only a mathematical
 idealisation never realised in the wavefunction of any actual
 physical system. 
 \par In conclusion, some brief remarks are made on some widely-held conceptions
  concerning the meaning of the EPR paper, in the light of the above considerations.
  \par \underline{EPR showed that quantum mechanics is `incomplete'}
   \par They did not. Correcting their logical errors and replacing `$TRUE$' on the right
       side of the proposition (4) by `$FALSE$' it might be concluded that it was 
       shown that quantum mechanics is `complete'. However, as explained above, the
       gedanken experiment cannot be performed in the real world, essentially 
       because of quantum mechanical uncertainty. `Completeness' or `incompleteness'
       according to EPR's definition cannot therefore be established by consideration
        of any gedanken experiment, no matter how idealised.
    \par \underline{The `EPR Paradox'}
    \par Because EPR's interpretation of their gedanken experiment suggests that, since 
          both the position and momentum of a particle (albeit measured in different
          experiments) may both be exactly known, there must be a contradiction
          with the momentum-space uncertainty relation. Since the analysis 
          of the gedanken experiment is based only on quantum mechanics the 
         latter must then be self-contradictory and therefore wrong. This erroneous conclusion
         arises from misinterpretation of uncertainty relations when {\it a priori}
         information is derived from state preparation.
\par \underline{`Spooky action-at-a-distance' in quantum mechanics}
  \par This problem results from the attempt to interpret the world governed by
       the laws of quantum mechanics in terms of classical concepts. The result of
       a quantum measurement at point A is no more `caused' by  another 
       one at a spatially separated point B than the measurement at B is `caused' by
       the one at A. Quantum mechanics predicts, however, that measurements
       at such causally-disconnected points may be correlated. Because of the
       principle of amplitude superposition in space-time\footnote{As, for example, in the
        Young double-slit experiment.} (a subject not  discussed in
      the EPR paper) quantum mechanical predictions are fundamentally
       non-local. That this application of classical causal concepts was
      stressed in the EPR paper by the introduction of 
      hypothetical `non-interacting' sub-systems is strongly related to Einstein's
      known and deep philosophical prejudice\footnote{See particlarly the
       passage from a letter from Einstein to Born in 1949 cited in Ref.~\cite{PT}. 
       According to the author of  the latter, the original source of Einstein's
       locality precept was Schopenhauer's {\it principium individuationis}.}
       concerning locality (the absence of action-at-a-distance) in physics~\cite{PT}.
       This prejudice is at variance with recent theoretical~\cite{JHFPS2} and experimental
       ~\cite{KSRM} indications for the non-retarded nature of electrodynamical force
       fields. For further discussion of this point see Ref.~\cite{JHFNREMF}.
       Almost at the end of Ref.~\cite{x1} occurs the following crucial 
        passage:
       \par `... We are thus forced to conclude that the quantum mecanical
         description of physical reality given by wave functions is not complete.
        \par One could object to this conclusion on the grounds that our
            criterion of reality is not sufficiently restrictive. Indeed, one would
     not arrive at our conclusion if one insisted that two or more physical
        quantities can be regarded as elements of reality {\it only when they can
         be simultaneously measured or predicted}. On this point of
         view, since either one or the other, but not both simultaneously,
        of the quantities $P$ and $Q$ can be predicted, they are not
        simultaneously real. \underline{This makes the reality of $P$ and $Q$
         depend upon the process of measurement}\newline \underline{ carried out on the
     first system which, does not disturb the second system 
         in} \newline \underline{any way.} No reasonable definition of reality could be expected 
         to permit this.'
        \par (Italics in the original)
        \par Notice the hypothetical causally disconnected subsystems that are 
             invoked in the underlined sentence. Such subsystems do not occur
             in the quantum mechanical description of nature.

{\bf Acknowledgements}
\par I thank N.Gisin and D.J.Moore for reading an early version of this paper and for
 their critical comments. Thanks are also due to two anonymous referees, of a journal which
 subsequently rejected this paper for publication;
 to the first, for pointing out that (4) is an `exclusive or' proposition, and
 to both of them for stressing that EPR's reasoning was presented as 
 a {\it reductio ad absurdum} argument.  

\pagebreak

\pagebreak 
\end{document}